\documentclass[twocolumn,showpacs,preprintnumbers,amsmath,amssymb]{revtex4}

\usepackage{graphicx}
\usepackage{dcolumn}
\usepackage{bm}

\begin{document}

\title{
The same physics underlying  SGRs, AXPs, and radio pulsars }
\author{Biping Gong}

\affiliation{ Department of Astronomy, Nanjing University, Nanjing
210093,PR.China}

\email{bpgong@nju.edu.cn}
\begin{abstract}

Unexpected sign, significant  magnitude and time-variation  of
frequency second derivative exist not only in singular radio
pulsars but also in Soft Gamma repeaters (SGRs) and Anomalous
X-ray pulsars (AXPs). This paper shows that the these phenomena
are related, and  can be interpreted by a simple unified model,
long-term orbital effect. Thus many of previous ``isolated''
pulsars may be binary pulsars, i.e., orbital period
 $P_b\approx(47, 72)$min for AXP 1E
2259+586, and $P_b\approx (20, 34)$min for PSR J1614-5047,
$P_b\approx (3.6, 6.4)$min for SGR 1900+14, and $P_b\approx (1.5,
5.8)$min for SGR 1806-20. In addition to X-ray pulsar, 1E
1207.4-5209, these two SGRs may be the new  ultra-compact binary
pulsars with orbital period of a few minutes. There might be more
binary pulsars that suitable for gravitational wave detection than
we had thought.

\end{abstract}
\draft 

\maketitle

\section{Introduction}
Pulsars are powered by rotational kinetic energy and lose energy
by accelerating particle winds and by emitting electromagnetic
radiation at their rotational frequency, $\nu$. The slowdown is
usually described by
\begin{equation}
\label{dipo1} \dot{\nu}=-\kappa{\nu}^n \ ,
\end{equation}
where $\kappa$ is a positive constant which determined  by the
moment of inertia and the magnetic dipole moment of the pulsar and
$n$ is the braking index. By Eq.($\ref{dipo1}$) we have,
\begin{equation}
\label{dipo2} \ddot{\nu}/\dot{\nu}=n\dot{\nu}/\nu \ ,
\end{equation}
$n=3$ for constant spin-dipole angle and dipole moment. Distortion
of the magnetic field lines in the radial direction from that of a
pure dipole, pulsar wind, and time-variable effective magnetic
moment results in $1\leq n\leq 3$~\cite{Dick85, br88}.

However a large numbers of  first and second pulse derivatives of
pulsars~\cite{Hobbs04} shows that majority pulsars differ from
$n=3$ significantly,
$$|\ddot{\nu}_{obs}/\dot{\nu}_{obs}|\gg|\ddot{\nu}/\dot{\nu}| =3|\dot{\nu}/\nu| \
.$$
Most frequency second derivative reported in the literature
deviate significantly from the simple dipole braking, and is
usually treated as long-term timing noise. The main
characteristics of the frequency second derivative are: (1) the
magnitude of it depends on the length of the data span; (2) the
sign of it can be both positive and negative; (3) the magnitude of
$\ddot{\nu}_{obs}$ can be orders of magnitude larger than that
expected by magnetic dipole radiation.

Gong~cite{Gong05a} introduced long-term orbital effect to explain
the puzzles of radio quiet neutron star 1E 1207.4-5209, in which
the discrepancy between the measured $\dot{\nu}_{obs}$ and the
magnetic dipole radiation induced $\dot{\nu}$ is attributed to the
long-term orbital effect.  This model actually provides a
mechanism that can interpret the three characteristics of timing
noise.

This paper extend the model in three aspects firstly to different
kinds of pulsars, SGRs, AXPs and radio pulsars; secondly to higher
order of frequency derivatives, which can explain puzzles in
braking index and timing noise, and finally the orbital period of
5 different pulsars are estimated which can put the model under
extensive test.

\section{The model}
The time for the pulsed light to travel across the projection of
the orbit into observer's line of sight from the instantaneous
position of the pulsar is
\begin{equation}
\label{delay} \frac{z}{c}=\frac{r\sin i}{c}\sin(\omega+f) \,,
\end{equation}
where $c$ is speed of light. $r$ is the distance between the focus
and the pulsar, $f$ is the true anomaly, $\omega$ is the angular
distance of periastron from the node, $i$ is orbital inclination,
as shown in Fig~1. The second of the phenomena due to orbital
motion is the change of pulse frequency, $\Delta\nu$,
\begin{equation}
\label{vnc}\frac{\Delta\nu}{\nu}=\frac{{\bf v}\cdot{\bf
n}_p}{c}=K[\cos(\omega+f)+e\cos\omega] \,,
\end{equation}
where $K\equiv {2\pi a_p\sin i}/{[cP_b(1-e^2)^{1/2}]}$  is the
semi-amplitude,
 $e$,
$P_b$, $a_p$ are eccentricity, orbital period, and pulsar
semi-major axis respectively.


What if a pulsar is in a binary system, however the effect of
Roemer time delay and Doppler shift, as given in
Eq.($\ref{delay}$) and Eq.($\ref{vnc}$) respectively, has not been
measured? In such circumstance what the observer measured is
neither as a true isolated pulsar, no as an usual binary pulsar
(which has measured Roemer delay or Doppler shift directly).

When a pulsar has a companion,  the time received by the observer
(Baryon centric time) is,
\begin{equation}
\label{tb} t_b=t_p+\frac{z}{c} \,,
\end{equation}
where $t_p$ is the proper time of the pulsar, and $z/c$ is
dependent of Kepler equation,
\begin{equation} \label{EM} E-e\sin
E=\bar{M}=\bar{n}t  \,,
\end{equation}
where $\bar{M}$, $E$ and $\bar{n}$ are mean anomaly, eccentric
anomaly and mean angular velocity respectively. Notice that $t$ is
the time of periastron passage, which is  uniform.

Obviously for a true isolated pulsar, we have $z/c=0$ in
Eq.($\ref{tb}$), thus $t_b=t_p$, which means both $t_b$ and $t_p$
are uniform. But for a binary pulsar system, $t_b$ is no longer
uniform, whereas  $t_p$ is still uniform.

Therefore, the proper time of the pulsar, $t_p$, can be used to
replace the uniform time, $t$ of Eq.($\ref{EM}$), then we have
$\bar{M}=\bar{n}t_p$.

If  $\Delta\nu$ of  Eq.($\ref{vnc}$) is averaged  over one orbit
period by the measured time, $t_b$, then it gives
$$
<\Delta\nu>=\frac{1}{P_b}\int^{P_b}_{0}\Delta\nu
dt_b=\frac{1}{P_b}\int^{P_b}_{0}\Delta\nu(dt_p+\frac{\dot{z}}{c}dt_p)$$
$$
=\frac{1}{P_b}\int^{P_b}_{0}\Delta\nu\frac{\dot{z}}{c}dt_p=\frac{x}{P_b}\int^{P_b}_{0}\Delta\nu
\cos(\omega+E)\dot{E}dt_p
$$
$$
=\frac{xK\nu}{P_b}\int^{2\pi}_{0}[\cos(\omega+f)+e\cos\omega]
\cos(\omega+E)dE $$
\begin{equation}\label{av1}
=\frac{xK\nu}{P_b}\pi(1-\frac{e^2}{4})+O(e^4)  \,,
\end{equation}
where $x$ is the projected semi-major axis, $x\equiv a_p\sin i/c$.

In practical observation, an observer may average $\Delta\nu$ from
0 to $T$ ($T\gg P_b$) through $t_b$, the time received by
observer, without knowing the orbital period, $P_b$, at all.
However if the pulsar measured is truly in a binary system, $P_b$,
will affect the averaged result, as given in Eq.($\ref{av1}$),
thus the averaged $\Delta\nu$ given by observer is,
$$ <\Delta\nu>= \frac{1}{T}\int^{T}_{0}\Delta\nu
dt_b=\frac{1}{T}[\int^{P_b}_{0}\Delta\nu dt_b+...+$$
\begin{equation} \label{av2}
\int^{P_b N}_{P_b (N-1)}\Delta\nu dt_b+\int^{T}_{NP_b}\Delta\nu
dt_b]=\beta+ o(\beta\frac{P_b}{T})
 \ ,
\end{equation}
where $\beta\equiv\pi(1-{e^2}/{4}){xK\nu}/{P_b}$. Eq.($\ref{av1}$)
and Eq.($\ref{av2}$) indicate that if a pulsar is in a binary
system, then $\Delta\nu$ (for convenience brackets, $<,>$ are not
used hereafter) measured by the observer is actually contaminated
by orbital effect, $\beta$. Comparatively, for a truly isolated
pulsar, there is no orbital effect, and thus $\beta=0$. And for a
pulsar that has already been recognized as in a binary system, the
effect of $\beta$ has been absorbed by binary parameters, such as,
$P_b$, $e$, and $\dot{\omega}_{GR}$,  the well known advance of
periastron given by General Relativity.

The derivative of $\Delta\nu$ can be obtained from
Eq.($\ref{av1}$) and Eq.($\ref{av2}$)~\cite{Gong05a},
\begin{equation}
\label{eq2}
\dot{\nu}_{L}=\beta\frac{\dot{a}}{a}(1-\xi)
 \,,
\end{equation}
where $a$ is the semi major axis of the orbit, and
$\xi\equiv\frac{(1-e^2)e^2}{2(1+e^2)(1-e^2/4)}+\frac{e^2}{1+e^2}$.
By the expression of $\dot{a}/a$~\cite{Gong05b}, it is related to
the advance of precession of periastron, $\dot{\omega}_{GR}$ by
\begin{equation}
\label{aaom}\frac{\dot{a}/{a}}{\dot{\omega}_{GR}}=\frac{M_1M_2}{3M^2}\frac{1+e^2}{(1-e^2)^{3/2}}(2+\frac{3M_2}{2M_1})(P_yQ_x-P_xQ_y)
 \,,
\end{equation}
where $M_1$, and $M_2$ are the mass of the pulsar, companion,
respectively  and $M$ is the  total mass of a binary system.
$P_x$, $P_y$, $Q_x$, $Q_y$ are sine and cosine functions of
$\omega$ (the angle distance of periastron from the node) and
$\Omega$ (the longitude of the ascending node)~\cite{Gong05b}.
Notice that $\dot{a}/{a}$ is a long periodic term, whereas,
$\dot{\omega}_{GR}$ is a secular term.

Considering Eq.($\ref{eq2}$), the observational $\dot{\nu}_{obs}$
is given
\begin{equation}
\label{dotpobs} \dot{\nu}_{obs}=\dot{\nu}+\dot{\nu}_{L}  \ ,
\end{equation}
where $\dot{\nu}$ is the intrinsic one, which caused by magnetic
dipole radiation and  $\dot{\nu}_L$ is caused by orbital effect.

\section{Application to SGRs}
SGRs and AXPs are believed to be magnetars. Like radio pulsars
these sources are spinning down and show timing noise, however at
magnitude on average 100 times larger.

 The derivatives of SGR 1900+14 satisfies following  relation,
\begin{equation}
\label{disc}|\frac{\ddot{\nu}_{obs}}{\dot{\nu}_{obs}}|
\sim|\frac{{\nu}^{(3)}_{obs}}{\ddot{\nu}_{obs}}|\sim|\frac{{\nu}^{(4)}_{obs}}{{\nu}^{(3)}_{obs}}|
\ .
\end{equation}
The derivatives of SGR 1806-20 satisfies following  relation,
\begin{equation}
\label{disc2}|\frac{\ddot{\nu}_{obs}}{\dot{\nu}_{obs}}|\gg
|\frac{{\nu}^{(3)}_{obs}}{\ddot{\nu}_{obs}}|\sim|\frac{{\nu}^{(4)}_{obs}}{{\nu}^{(3)}_{obs}}|
\ .
\end{equation}
The magnetic dipole radiation satisfies
\begin{equation}
\label{disc3} \ddot{\nu}/\dot{\nu}=3\dot{\nu}/\nu \ , \ \
{\nu}^{(3)}/\ddot{\nu}=5\dot{\nu}/\nu \ , \ \
{\nu}^{(4)}/{\nu}^{(3)}=7\dot{\nu}/\nu \,.
\end{equation}
Thus it seems that the relation, Eq.($\ref{disc}$), can be
explained by magnetic dipole radiation. However the magnitude of
$\ddot{\nu}/\dot{\nu}$ is  much smaller than observational ratio
given by Eq.($\ref{disc}$). Therefore, the relation
Eq.($\ref{disc}$) cannot be explained by magnetic dipole
radiation, neither is Eq.($\ref{disc2}$).

Through Eq.($\ref{dotpobs}$) following relation can be obtained,
\begin{equation}
\label{dipole4}
\frac{\ddot{\nu}_{obs}}{\dot{\nu}_{obs}}=\frac{3\dot{\nu}}{\nu}\frac{\dot{\nu}}{\dot{\nu}_{obs}}
+\dot{\omega}_1\frac{\dot{\nu}_{L}}{\dot{\nu}_{obs}} \ \,,
\end{equation}
where $\dot{\omega}_1\equiv{\ddot{\nu}_{L}}/{\dot{\nu}_{L}}$. The
first term at the right hand side of  Eq.($\ref{dipole4}$)
corresponds to the magnetic dipole radiation and the second one
corresponds to the orbital effect which can change sign and have
much larger magnitude than that of the first term at right hand
side of  Eq.($\ref{dipole4}$). This actually explains why  $|n|\gg
3$ is inevitable when the second term is ignored.

Similarly by Eq.($\ref{dotpobs}$) we can have,
\begin{equation}
\label{disc4} \frac{{\nu}^{(3)}_{obs}}{\ddot{\nu}_{obs}}=
\frac{5\dot{\nu}}{\nu}\frac{\ddot{\nu}}{\ddot{\nu}_{obs}}
+\dot{\omega}_2\frac{\ddot{\nu}_{L}}{\ddot{\nu}_{obs}} \ \,,
\end{equation}
where $\dot{\omega}_{2}\equiv{\nu}^{(3)}_{L}/\ddot{\nu}_{L}$.
Again the left hand side of Eq.($\ref{disc4}$) can be explained by
the second term at the right hand side which is dominant.
$\dot{\omega}_{1}$ and $\dot{\omega}_{2}$ are generally in order
of magnitude, $\dot{\omega}$ or $\dot{\Omega}$. However since they
are all long periodic terms  which dependent of $\omega$ and
$\Omega$, at certain time the trigonometric function of $\omega$
and $\Omega$ may cause a relative large or small values in the
derivative or derivatives of pulse frequency, and  results the
discrepancy between Eq.($\ref{disc}$) and Eq.($\ref{disc2}$).


The magnetic dipole radiation induced frequency first derivative
is always negative; whereas, the long-term orbital effect can
cause both negative and positive $\dot{\nu}_{L}$. The fact that
most pulsars have negative frequency derivative indicates that
$|\dot{\nu}|>|\dot{\nu}_{L}|$. In other words, for most pulsars
$\dot{\nu}_{obs}$ is dominated by $\dot{\nu}$, thus
$\dot{\nu}_{obs}$ and $\dot{\nu}$ have the same sign. Therefore we
can assume: $\dot{\nu}=\sigma\dot{\nu}_{obs}$, and in turn
$\dot{\nu}_{L}=(1-\sigma)\dot{\nu}_{obs}$, where $\sigma>0.5$. For
convenience define $\alpha=1-\sigma$, notice that $\alpha$ can
both be positive and negative. Eq.($\ref{dipole4}$) can be
rewritten,
\begin{equation}
\label{eq1}
\frac{\ddot{\nu}_{obs}}{\dot{\nu}_{obs}}=\frac{3(1-\alpha)^2\dot{\nu}_{obs}}{\nu}
+\dot{\omega}_1\alpha\ \,,
\end{equation}
Actually $\alpha$ can be obtained from observation, from which
$\dot{\omega}_1$ of Eq.($\ref{dipole4}$) can be obtained.  However
$\dot{\omega}_1$ may deviate from $\dot{\omega}_{GR}$,
\begin{equation}
\label{omom} \dot{\omega}_{GR}=\gamma\dot{\omega}_1 \ \,,
\end{equation}
By assuming  $\gamma$ (typically  $\sim10$) one can obtain
$\dot{\omega}_{GR}$, from which the orbital period can be
estimated,
\begin{equation}
\label{dipole5} P_b=2\pi[\frac{3(
GM)^{2/3}}{c^2(1-e^2)\dot{\omega}_{GR}}]^{3/5}\ \,,
\end{equation}
where  $G$ is the gravitational constant. Having $P_b$, the
semi-major axis of orbit, $a$ of Eq.($\ref{eq23}$),  can be
obtained. Finally putting $\dot{\omega}_1$, $a$, $P_b$, as well as
estimated  $M_2$ ($M_1=1.4M_{\odot}$) into following equation
(which is given by rewritten Eq.($\ref{eq2}$)),
\begin{equation}
\label{eq23}
\alpha\dot{\nu}_{obs}=\dot{\nu}_{L}=\frac{GM\nu}{2\pi c^2
a}\rho\frac{\dot{a}}{a}(1-\xi)=\frac{GM\nu}{2\pi c^2
a}\rho\dot{\omega}_1
 \,,
\end{equation}
where $\rho\equiv\pi\sin^2 i(M_2/M)^2(1-e^2/4)/\sqrt{1-e^2}$, we
can  obtain $\rho$. Then one can adjust the  companion mass,
$M_2$,  and check whether $|\sin i|\leq 1$ is satisfied or not in
the expression of $\rho$. Moreover putting $M_2$, $\sin i$ and
$P_b$ into
 Eq.($\ref{av2}$) one can obtain $\Delta\nu/\nu$ induced by the
orbital effect, and further check whether it can be consistent
with $\Delta\nu/\nu$ given by observation or not.

In other words, one can adjust two parameters, $\gamma$ and $M_2$
to satisfy the three constraints, $\alpha$, $\sin i$ and
$\Delta\nu/\nu$.

By the observation of SGR 1900+14,  $\dot{P}_{obs}=8.2(6)\times
10^{-11}$s~s$^{-1}$ in May 31-Jun 9, 1998; and
$\dot{P}_{obs}=5.93(3)\times 10^{-11}$s~s$^{-1}$ in Aug 28-Oct 8,
1999~\cite{Woods99}, $\dot{\nu}_{obs}$ varies significantly
($\nu_{obs}=1/P_{obs}$). By $\alpha=0.32$, and through
Eq.($\ref{eq1}$) and Eq.($\ref{dipole5}$), we can obtain
$P_b\approx 6.4$min. By assuming  $M_1=1.4M_{\odot}$,
$M_2=1.4M_{\odot}$ $\gamma=10$, we have $\sin i=0.2$ and $x=0.04$
through the definition of $\rho$. Another solution is $P_b\approx
3.6$min, $M_1=1.4M_{\odot}$, $M_2=0.5M_{\odot}$ and $\gamma=20$ as
shown in Table~1.

Nevertheless $|\alpha|<0.5$ guarantees that the  magnitude of the
frequency first derivative caused by the long-term orbital effect
is smaller than that of the intrinsic spin down. This means the
long-term orbital effect model is not contradictory to the
assumption that SGRs have high magnetic field, while it is not
contradictory to other possibilities either~\cite{Marsden99,
Mosq04}.

On the other hand,   the  second order frequency derivatives
caused by the long-term orbital effect is much larger than that of
the magnetic dipole radiation induced one for SGRs,  in other
words,  $\ddot{\nu}_{obs}$ is dominated by $\ddot{\nu}_L$. And
since $\ddot{\nu}_L$ can change sign at  time scale
$\sim2\pi/\dot{\omega}_{GR}$; it is expected that
$\ddot{\nu}_{obs}$ (also  ${\nu}^{(3)}_{obs}$,
${\nu}^{(4)}_{obs}$) will change sign at  time scale
$\sim2\pi/\dot{\omega}_{GR}$. This can be tested by observation.

\section{Application to AXPs and radio pulsars}
The variation of $\dot{\nu}_{obs}$ and $\ddot{\nu}_{obs}$  between
1990 January and 1998 December of 11 pulsars is measured using
ATNF Parkes radio telescope~\cite{Wang00}. The signs of
$\ddot{\nu}_{obs}$ of  PSR J1614-5047 and PSR 1341-6220 change for
two times, which have been attributed to glitch.

The change on $\dot{\nu}_{obs}$ of PSR J1614-5047 is about $1\%$,
thus  $\alpha=0.01$, and similarly through
Eq.($\ref{dipole4}$)--Eq.($\ref{dipole5}$), the orbital period of
PSR J1614-5047 can be estimated, $P_b\approx 34$min. This result
corresponds to $M_2=0.7M_{\odot}$, $\sin i=0.2$ and $x=0.05$s as
shown in Table~1.

 The measured time scale of change sign on
$\ddot{\nu}_{obs}$ of PSR J1614-5047 is $\sim3.2$yr, which is
consistent $2\pi/\dot{\omega}_{GR}\sim3$ yr corresponding to
$P_b\approx 34$min.


By Eq.($\ref{av2}$) and Eq.($\ref{eq2}$), $\Delta\nu$ and
$\dot{\nu}_{L}$  contains orbital elements, $a$, $e$, and $i$,
which are all long-periodic under the Spin-Orbit coupling model.
Thus $\Delta\nu$ and $\dot{\nu}_{L}$ are trigonometric functions
of $\omega$ and $\Omega$ (Gong 2005a). In such case $\omega$ and
$\Omega$ are non-uniform which may induce abrupt change in
$\Delta\nu$ and $\dot{\nu}_{L}$, and therefore mimic glitches in
pulsars. The absence of glitches in binary pulsars may be due to
that the additional times delay caused by long-term orbital effect
can be absorbed by the uncertainties in parameters such as $P_b$,
$\dot{\omega}_{GR}$ and $\dot{e}$. Whereas, for isolated pulsars,
the only possible absorbtion of long-term orbital effect is by
rotational parameters, ${\nu}_{obs}$ and $\dot{\nu}_{obs}$, etc.
This explains why glitches always happens in young isolated
pulsars. As shown in Table~1, these pulsars usually have short
orbital period, i.e., from a few minute to $10^1$min, which can
make significant timing noise, but the small semi-major axis,
$x\sim 10^{-2}$ to $10^{-3}$, as shown in Table~1,  prevents them
from being observed as binary pulsars directly.

The timing noise parameter is defined as $\Delta(t)\equiv
log(|\ddot{\nu}|t^3)/6\nu$, and $\Delta_8$ means $\Delta(t=10^8)$.
By Eq.($\ref{eq23}$), the second derivatives of ${\nu}_{L}$ is
given,
\begin{equation}
\label{dnu2} \ddot{\nu}_{L}\approx\frac{GM\nu}{2\pi c^2
a}\rho\ddot{\omega}_1
 \,,
\end{equation}
Eq.($\ref{dnu2}$) indicates that $\ddot{\nu}_{L}\propto \rho$, and
by the definition of $\rho$, we have $\ddot{\nu}_{L}\propto \sin^2
i , M_2^2 , 1/P_b^{2/3}$. This explains why PSR J1314-6220 has the
minimum timing noise, $\Delta_8\approx 0.0012$, and   SGR 1806-20
has the maximum, $\Delta_8=5.5$ among the five pulsars shown in
Table~1.

\section{Discussion}


Different timing behaviors shown in different kinds of pulsars may
caused by the same physics. The long-term orbital effect provides
possible explanations to following phenomena:

1. Magnitude and time scale on the variation of pulse frequency,
$\Delta\nu$.

2. Magnitude, sign and variation of frequency second derivative,
${\ddot{\nu}_{obs}}$.

3. The relationship of ratios of derivatives of pulse frequency
like ${\ddot{\nu}_{obs}}/{\dot{\nu}_{obs}}$,
${{\nu}^{(3)}_{obs}}/{\ddot{\nu}_{obs}}$.

4. Why unexpected sign of $\ddot{\nu}_{obs}$ appears much more
often than that of  $\dot{\nu}_{obs}$.

5. Why  timing noise parameters, $\Delta_8$, of SGRs are much
larger than that of radio pulsars.

There are two predictions from to the new model.
$\ddot{\nu}_{obs}$, ${\nu}^{(3)}_{obs}$ and ${\nu}^{(4)}_{obs}$ of
SGRs, AXPs and radio pulsars  should change sign at time scale
$\sim 2\pi/\dot{\omega}_{GR}$. The test of this prediction may
tell us whether timing noise is  caused by the long term orbital
effect or not.

Another prediction is orbital period of different pulsars listed
in Table~1. The two SGRs may be ultra-compact binary with orbital
period of a few minutes, this is a natural extension from the
binaries with orbital period of $10^1$min. If confirmed the
population of sources for gravitational wave detectors, like LIGO
and LISA, may increase considerably.

\section{Acknowledgments}

I would like  to thank T.Y. Huang and X.S. Wan for helpful
discussion and comment on the presentation of the manuscript.
I also thank  W.D. Ni, X.D. Li, X.L. Luo, Q.H. Peng, Z.G. Dai,
R.X. Xu for use helpful discussion.

\clearpage
\begin{table}
\begin{center}
\caption{Calculated rotational parameters and estimated orbital
parameters of possible binary pulsars}
\begin{tabular}{crrrr|rrrrrr}
\hline \hline

pulsars & $n$ &  $\Delta_8$  & $\dot{\nu}_L/\dot{\nu}_{obs}$  &
refs & $P_b$(min) & $m_2(M_{\odot})$ & $\sin i$ & $x$ & $\gamma$ &
e

\\\hline

1E 2259+586 & 3302 & $0.42$ & 0.10  & 1 & 72 & 0.1 & 0.6 & 0.06 & 10 & 0.2\\
 & & & & &  47 & 0.1 & 0.4 & 0.03 & 20 & 0.2\\
SGR 1806$-$20  & -826 & 5.5 & 0.32 & 2 &  5.8 & 0.5 & 0.4 & 0.03 & 1 & 0.2$^a$\\

 & & & & &  1.5 & 0.1 & 0.4 & 0.03 & 1 & 0.2$^b$\\

SGR 1900+14 & -3921 & 4.5 & 0.30 & 3 &  6.4 & 1.4 & 0.2 & 0.04 & 10 & 0.1\\

 & & & & &  3.6 & 0.5 & 0.2 & 0.01 & 20 & 0.1 \\

PSR J1341$-$6220 & -1 & $0.032$ & 0.0012 & 4 &   27 & 0.01 & 0.2 & 0.001 & 10 & 0.1\\

 & & & & &  46 & 0.02 & 0.3 & 0.004 & 4 & 0.1\\

PSR J1614$-$5047 & -81 & $0.12$& 0.010 & 4 &   20 & 0.1 & 0.2 & 0.01 & 10 & 0.1\\
 & & & & &  34 & 0.7 & 0.2 & 0.05 & 5 & 0.1\\

\hline \hline
\end{tabular}
\end{center}
{\small $^a$ and $^b$ correspond to observational data of 1999 and
2000 respectively. Notice that the orbital parameters of SGR
1806$-$20 are estimated through $\dot{\omega}_2$ of
Eq.($\ref{disc4}$), instead of $\dot{\omega}_1$ of
Eq.($\ref{dipole4}$) as the rest of pulsars in this table.
$\dot{\omega}_2$ is very  close to $\dot{\omega}_{GR}$, which
corresponds to $\gamma\approx1$. The test of the binary nature of
these pulsars can be performed by setting
$\dot{\nu}_{obs}=\dot{\nu}$, and $\ddot{\nu}_{obs}=\ddot{\nu}$, in
other words, $\dot{\nu}_{obs}$ and $\ddot{\nu}_{obs}$ satisfies
the expectation of magnetic dipole radiation, and then use orbital
parameters to fit the quasi-sinusoidal residual. Orbital  period,
$P_b$, predicted in this table is obtained in the case that the
mass of the pulsar is $M_1=1.4M_{\odot}$. The error of the
estimated $P_b$ mainly comes from the assumption
$\dot{\omega}_{GR}=\gamma\dot{\omega}_1$, typically the error in
$\gamma$ is less than 10, thus by Eq.($\ref{dipole5}$), the error
of $P_b$ is $P^{\ +\sigma 1}_{b~-\sigma 2}$ (where $\sigma
1\equiv{3P_b}$ and $\sigma 2\equiv{3P_b/4}$). $n$, $\Delta_8$ and
$\dot{\nu}_L/\dot{\nu}_{obs}$ are obtained by references: 1 Kaspi
et al~\cite{Kaspi03}; 2 Woods et al\cite{Woods00}; 3 Woods et
al\cite{Woods02}; 4 Wang et al\cite{Wang00}. }
\end{table}


\begin{thebibliography}{99}
%

\bibitem{Dick85} Manchester, R.N., Durdin, J.M., $\&$ Newton, L.M., 1985, Nature
313, 374.

\bibitem{br88} Blandford, R.D. $\&$ Romani,
R.W. 1988, MNRAS, 234, 57.

\bibitem{Hobbs04} Hobbs, G., Lyne, A. G., Kramer, M.,
Martin, C.E., $\&$ Jordan, C. 2004, MNRAS, 353, 1311-1344.





\bibitem{Gong05a} Gong, B.P. Astro-ph/0506431, submitted.

\bibitem{Gong05b} Gong, B.P. Chin. J. Astron.  Astrophy.  in press (2005).


\bibitem{Woods99} Woods, P.M., Kouveliotou, C.,
Paradijs, J.V., Finger, M.H., Thompson,C., Duncan, R.C.,  Hurley,
K., Strohmayer, T., Swank, J., $\&$ Murakami, T. 1999, ApJ. 524,
L55-L58 .

\bibitem{Marsden99} Marsden, D., Rothschild, R.E. $\&$ Lingenfelter,R.E.,  1999. ApJ, 520, L107--L110.

\bibitem{Mosq04} Mosquera Cuesta, H.J., Astro-ph/0411513


\bibitem{Wang00} Wang, N., Manchester, R. N., Pace, R., Bailes, M., Kaspi, V. M.,
Stappers, B. W. $\&$ Lyne, A. G., 2000. MNRAS, 317, 843--860.





\bibitem{Kaspi03} Kaspi,V.M., Gavriil,F.P, Woods, P.M., Jensen, J.B.,
Roberts, M.S.E., $\&$  Chakrabarty,D. 2003,  ApJ. 588, L93-L96 .
%



%





\bibitem{Woods02} Woods, P.M., Kouveliotou, C., G\"og\"us, E.,
Finger, M.H.,  Swank, J., Markwardt, C.B., Hurley, K.,$\&$ Klis,
M.V. ApJ. 2002,  576, 381-390 .

\bibitem{Woods00} Woods, P.M., Kouveliotou, C.,
Finger, M.H., G\"og\"us, E., Scott, D.M., Dieters, S.,
Thompson,C., Duncan, R.C.,  Hurley, K., Strohmayer, T., Swank, J.,
$\&$ Murakami, T. 2000, ApJ. 535, L55-L58 .



\end{thebibliography}
\end{document}